# Internal motion of soft granular particles under circular shearing: Rate-dependent quaking and its spatial structure

Jr-Jiun Lin(林祉均)[1,2,*], Cheng-En Tsai(蔡承恩)[3,1], Jung-Ren Huang(黃仲仁) [1], and Jih-Chiang Tsai(蔡日強) [1,*]

[1] Institute of Physics, Academia Sinica, Taipei, Taiwan
[2] Physics Department, National Tsing Hua University, Hsin-Chu, Taiwan
[3] Physics Department, National Taiwan University, Taipei, Taiwan

Correspondence (*): Jr-Jiun Lin < linjrjun@gmail.com>, JC Tsai<jctsai@phys.sinica.edu.tw>

Tightly packed granular particles under shear often exhibit intriguing intermittencies, specifically, sudden stress drops that we refer to as quaking. To probe the nature of this phenomenon, we prototype a circular shear cell that is capable of imposing a uniform and unlimited shear strain under quasi-static cyclic driving. Spherical PDMS(polydimethylsiloxane) particles, immersed in fluid, are driven in a fixed total volume at a wide range of shear rates, with particle trajectories captured in 3D space via refraction-index-matched fluorescent tomography. Statistics on the magnitude of fluctuating displacements of individual particles show a distinct dependence on the shear rate. Particles move smoothly at high shear rates. At intermediate shear rates, quaking emerges with clusters of particles exhibiting relatively large displacements. At low shear rates, a cluster can span the entire system, and the cluster exhibits substructures in view of localized particle movements. Overall, we have confirmed that the quaking phenomena in the current setup are consistent with our previous work [*Phys. Rev. Lett.*,*126*, 128001 (2021)], and that the dimensionless shear rate that we have proposed [*Phys. Rev. Research 6*, 023065 (2024)] is indeed a good parameter for unifying the transitions observed in different experimental geometries.

# I. INTRODUCTION

Granular material exhibits multifaceted flow properties that are ubiquitous in nature, such as those directly related to phenomena ranging from surface creep [1–3] to earthquakes [4–6]. Such wide variety has motivated various physical descriptions under different conditions. In particular, there are two prevailing categories: visco-plastic models [7–10] and those derived from dense suspensions [11]. The former describes the regime dominated by interparticle contacts, in which the strain concentrates in highly localized plastic regions known as shear transformation zones [9,12,13]. Meanwhile, the latter invokes the viscous number and shear thickening behaviors [14-16] to capture scenarios where the interstitial fluid becomes dynamically relevant via lubrication and viscous force.

In an effort to bridge the two narratives, our previous experimental study [17] presents a continuous transition from "granular solid" to "granular fluid" that depends explicitly on shear rate. The onset of this solid-fluid transition is distinctly marked by the occurrence of intermittent stress releases, which are often referred to as "avalanche" or "quake" in the literature [18–21]. Such intermittency is asserted to stem from the speed dependence of inter-particle friction, a general tribological property vindicated by measurements [22,23]. In a further numerical study [24], we have shown that adding the speed dependence of interparticle friction is sufficient in generating the empirically observed quaking behaviors reported in Ref.[17]. The behaviors are well described by a dimensionless shear rate $S_\ell \equiv \ell \dot{\gamma}/V_c$, in which $\ell$ stands for particle size, $\dot{\gamma}$ the physical shear rate, and $V_c$ the material-specific speed beyond which the friction drops substantially. The parameter $S_\ell$ therefore makes the comparison of two speeds. The value of $\ell \dot{\gamma}$ specifies the rough magnitude of the speed at which particles slide against their neighbors. In the case $\ell \dot{\gamma} \gg V_c$ (so that $S_\ell \gg 1$), almost all contacting pairs are sliding against each other with negligible friction, while $\ell \dot{\gamma} \ll V_c$ (so that $S_\ell \ll 1$) creates the opposite scenario that all contacting pairs are in full Coulomb friction. While $S_\ell \gg 1$ and $S_\ell \ll 1$ stand for the scenarios with high lubrication and conventional Coulomb friction between particles, respectively, quaking prevails in granular flows with intermediate values of $S_\ell$ [24]. Note that the occurrence of stick-slip quaking marks the breakdown of the concept of viscous number ($J \equiv \eta \dot{\gamma}/P$) for the flow of granular particles that are sufficiently tight – see Ref.[17] for detailed explanations.

The grain-scale dynamics of granular flow have been widely explored via various experimental techniques. Previous works implement quasi-2D systems [4,21,25] that unveil individual grain motion as well as force chain structures. Meanwhile, 3D bulk imaging techniques, such as refraction index matched imaging [19,26,27], x-ray tomography [28–30], and magnetic resonance imaging [31,32] have been developed to capture the internal structure of granular media. These technologies enabled delicate 3D experiments that shed light on local response to shear before macroscopic yielding, such as the overall spatio-temporal correlation of local strain [19] and the network topological evolution [29,30].

In the present study, we take advantage of a fast 3D scanning scheme and capture the grain-scale kinematics of individual quaking events through a wide range of shear rates, all under the regime of low-inertia steady flow. We present the real-time 3D spatial structure of quaking events and its shear rate dependence, expanding the scope of our previous analysis [17] which mainly focuses on the transition of stress patterns and its implications. The main purpose of our work here is two-fold. We hope to demonstrate, experimentally, that the non-monotonic stick-slip behaviors with respect to the shear rate are generic, not specific to the double-cone geometry as reported in Ref.[17] and that both works can be connected by the dimensionless shear rate $S_\ell$. Secondly, we provide the grain-level observation of particle movements that we were unable to address in Ref.[17].

## II. SETUP AND OVERVIEW OF THE RATE-DEPENDENT TRANSITION

We have established a prototype of Circular Shear Cell (CSC), which is shown in Fig.1: The sidewall of the cell is formed by a stack of acrylic rings that are driven cyclically by four steel rods going through the rings. The four rods are rotating in phase, with their lower ends fixed at the base plate with a rotary joint. The orientation of these rods specifies a unit vector $\hat{s}$ that rotates at an adjustable angular speed $\Omega$ and defines three orthogonal unit vectors $\{\hat{s}, \hat{\phi}, \hat{z}\}$ at every instant as shown on the graph. In the laboratory frame of reference, the driving imposes a base flow $\overrightarrow{v^0}(\vec{x}, t) = \Omega (z - z_0) \tan \alpha \ \hat{\phi}$ in which $z_0$ stands for the height of the base plate and, at every instant, the flow field exhibits a spatial gradient that has a component $\frac{\partial v_\phi^0}{\partial z} \equiv \dot{\gamma}_{inst} = \Omega \tan \alpha$ while all other components are zero. The constant $\dot{\gamma}_{inst}$ can be regarded as an effective shear rate on the $\phi z$-plane. By design, we set the angle $\alpha$ to be sufficiently large ($\tan \alpha = 0.45$) so that the particles are incrementally rearranged and never come back to the same positions after one complete cycle of driving, as validated by particle trajectories detailed in Appendix A. (The effective shear strain for one complete driving cycle is $\gamma_0 \equiv \int \dot{\gamma}_{inst} dt = \int \Omega \tan \alpha \ dt = 2\pi \tan \alpha \approx 2.8$.)

The cell is packed at a fixed volume fraction = 0.59 with a binary mixture of molded PDMS (polydimethylsiloxane) spheres (Young's modulus $\approx 1.5$ MPa) having diameters 9 and 12 mm, respectively mixed at the number ratio 3:2. The CSC setup has a diameter $2R_{total} = 35$ cm and height $H_{total} = 10$ cm, which forms a bulk of $\approx 10^4$ particles. The interstitial space is filled with glycerol-water mixture that index-matches the particles containing LASER-activated fluorescent dye (Rhodamine B). Using a rotating cube, we are able to shift a horizontal LASER sheet (supplied by COHERENT Verdi G5, at 532nm) periodically by a distance up to 6 cm in the $z$-direction, at rates ranging from 10 to 0.008 Hz. Time-resolved tomographic images are captured via a high-speed camera (PHANTOM VEO) underneath. A sample image is shown in Fig. 2. Note that the two types of particles are dyed differently, allowing us to easily distinguish them in tomographic images. Due to multiple scattering, the LASER sheet disperses, resulting in blurred images at the father half of the imaging area. Therefore we adopt only half of the image that is closer to the LASER source for analyses --- see the boxed region in Fig. 2. The 3D image stack is formed by a series of 100 images captured over a complete scan. In our analyses, we consider a stack corresponding to a 12cm×20cm×6cm space containing roughly 1200 particles that are far from the boundary. The 3D trajectories of these particles can be obtained via TRACKPY [33,34]. The details of 3D image reconstruction and particle tracking can be found in Appendix B.

The base plate is constrained by four surrounding force sensors placed at an offset distance $r_{off}$ away from the central axis, as shown by the top view in Fig. 1. This configuration allows us to determine both the net force $\overrightarrow{f_{xy}}$ and the torque exerted on the base plate ---see Appendix C.

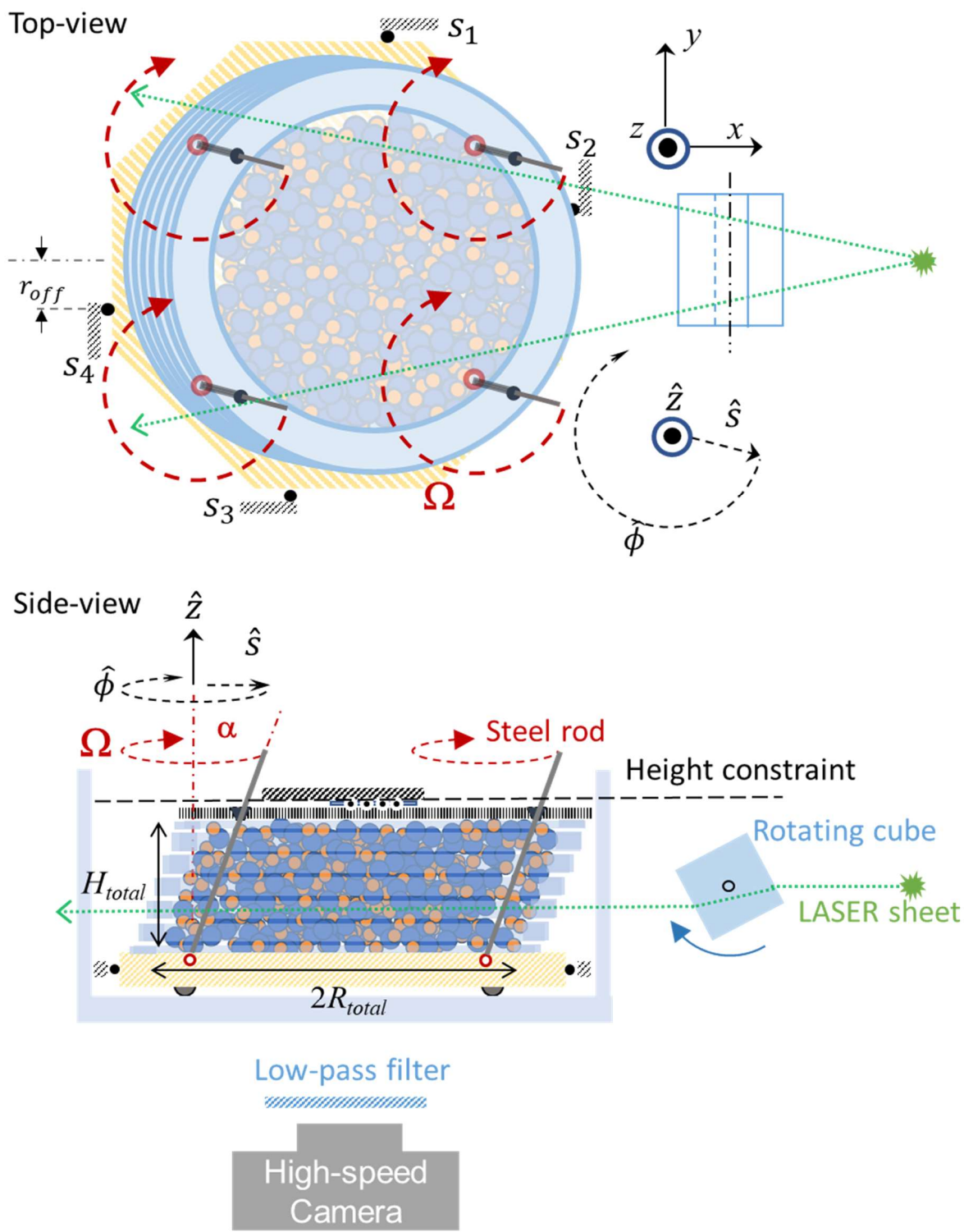

FIG.1 Schematics of the device: The top view highlights that the four steel rods (and in turn the acrylic rings) are doing circular motions sharing the same angular speed $\Omega$. At every instant, the orientation of the rods defines the unit vectors $\hat{s}$ and $\hat{\phi}$. The four force sensors that keep the octagonal base plate (yellow shaded) in place are labelled as $s_1 \sim s_4$. The side view shows that the rotating cube creates a periodic vertical scan by the LASER sheet that penetrates the granular packing, whose total volume is constrained from the top.

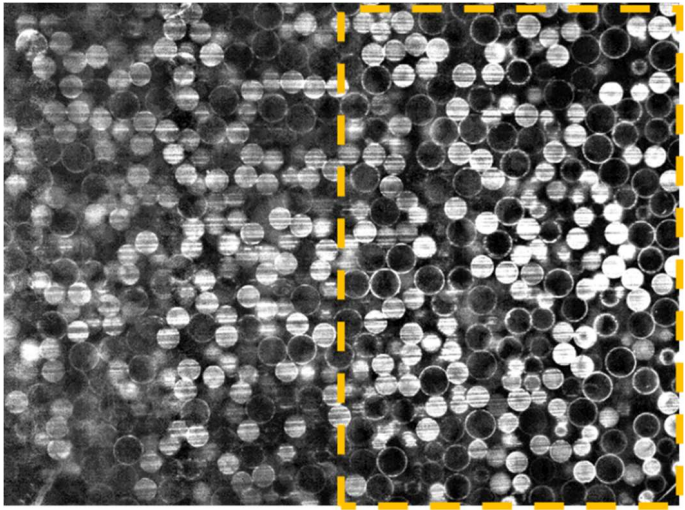

**FIG.2** –Typical tomographic image. The 9 mm particles appear as white disks, and the 12 mm particles as hollow circles. The yellow dashed box shows the half of the image used in our analyses.

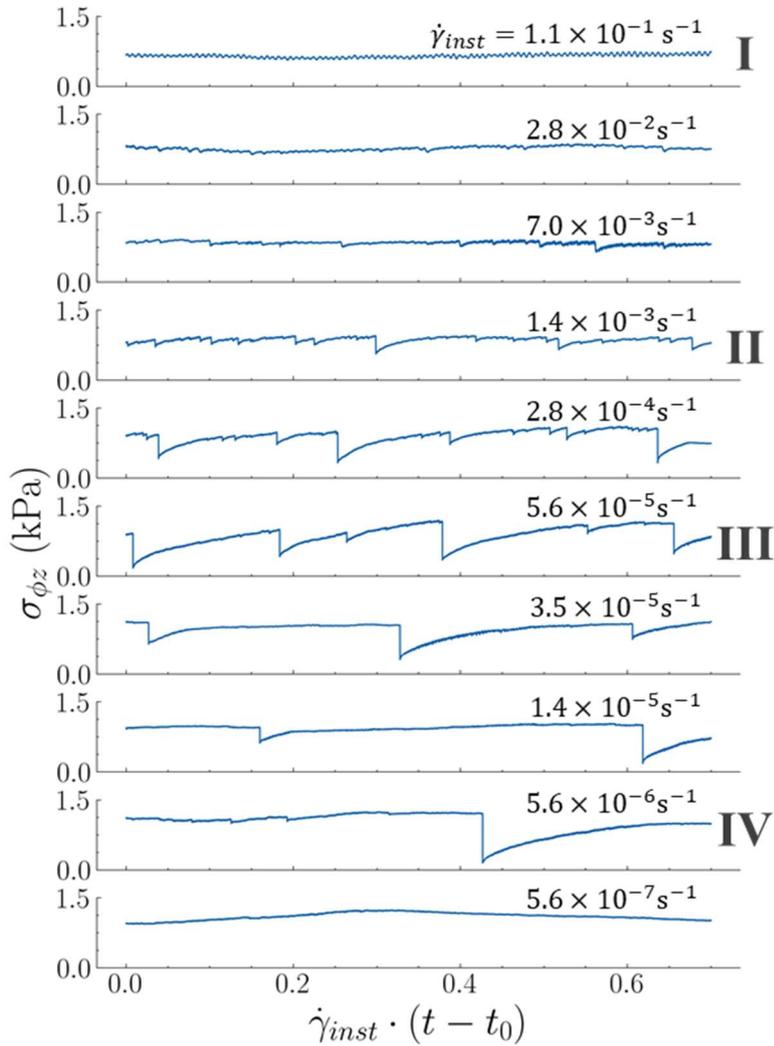

**FIG.3** Time sequence of the instantaneous shear stress $\sigma_{\phi z}$ for various values of shear rate (marked at the upper-right corner of every sub-plot). Roman numerals mark the four signature cases that will be referred to in the upcoming sections. For each run, we start the motor at $t = 0$, and all recordings begin at $t = t_0$ such that $\Omega t_0 = 0.5\pi$. This provides a preshear $\approx 0.7$ to ensure that the system has reached a steady state for each intended shear rate. We estimate that the inertial number is $7 \times 10^{-4}$ for the fastest case and $7 \times 10^{-9}$ for the slowest --- see main texts for the basis of our estimate.

The time sequences of the instantaneous shear stress $\sigma_{\phi z}$ in Fig.3 provide an overview of the non-monotonic dependence of the quaking behavior on the change of shear rate $\dot{\gamma}_{inst}$ across six orders of magnitude. The force that opposes the shear strain $f_{\phi} \equiv \overrightarrow{f_{xy}} \cdot (-\hat{\phi})$ is measured at sampling rates up to 1000 Hz. The instantaneous shear stress $\sigma_{\phi z}$ is determined as $f_{\phi}$ divided by the cross-sectional area of the cell ($961 \text{cm}^2$). In obtaining $f_{\phi}$, we have also removed the contribution from a mechanical artifact (around 10% of the mean value) --- please see Appendix C for its detail. For the convenience of further discussion, we use Roman numerals I~IV to indicate four signature cases on the figure. At high shear rates such as that of case I, $\sigma_{\phi z}$ is smooth with merely negligible high-frequency oscillations. As $\dot{\gamma}_{inst}$ decreases, the time sequence of shear stress displays increasingly larger sudden drops, which we refer to as "quakes". In case III, the drops reach magnitudes that are comparable to the mean value of $\sigma_{\phi z}$. Interestingly, as $\dot{\gamma}_{inst}$ further decreases, the system tends to develop large quakes that are separated by increasingly larger strain. Meanwhile, to make a fair comparison, we have kept the total effective strain fixed to be $\approx 0.7$ for all experiments. An accurate statistical count on these rare events at slow driving ratess can only rely on the repetition of long-time experiments and becomes increasingly difficult to obtain. Nevertheless, at the second lowest shear rate ($5.6 \times 10^{-6}$ s$^{-1}$) shown in Fig.3, we have repeated this experiment and do find one big event that is consistent with what is shown as case IV. We have further decreased the shear rate to $5.6 \times 10^{-7}$ s$^{-1}$ and found that, although the two weeks of experiment has exceeded the timespan for maintaining our index-matching (therefore we are unable to do particle tracking in this case), no quakes are detected within the effective strain $\approx 0.7$. This ensures us that the "statistical count" should be even lower than that of case IV. The time sequences of $\sigma_{\phi z}$ in Fig.3 thus provide a good overview of the non-monotonic dependence of the quaking behavior on the change of $\dot{\gamma}_{inst}$.

To obtain the inertial number $I = \rho^{0.5} d\dot{\gamma}/P^{0.5}$ in which $\rho$ stands for the mass density (=1030 Kg/m$^3$ for PDMS) and $d$ =0.01m, although we did not have a direct access the "confining pressure" $P$ at the time we performed these experiments (because we could only measure the shear stress, not the normal stress), we can still have a fair assessment on the inertial numbers involved: Using an improved setup (to be published in future work) at the same packing fraction, we are now able to determine the value of $\sigma_{zz}$ at a shear rate quite close to one of those reported here. Given that $I \propto \dot{\gamma}/P^{0.5}$ and that level of stress exhibits a mild increase over the decrease of shear rate (as shown in Fig.3), we estimate the inertial number in the experiments reported here ranges from $7 \times 10^{-4}$ for case I (the fastest) to $7 \times 10^{-9}$ for the slowest case. In other words, all cases reported here can be perceived as in the "quasi-static regime" of granular flow.

## III. MAIN RESULTS

We calculate the displacement $\Delta \vec{r}$ of each grain over one strain step $\Delta \gamma = \dot{\gamma}_{inst} \, \delta t$, which is set to be $\gamma_0/800 \approx 3.5 \times 10^{-3}$ for all runs except for two exceptions[1]. A generic snapshot of $\Delta \vec{r}$ vectors at a quiescent interval is shown in Fig. 4(a), which is 3D rendered via OVITO [35,36]. Particle movements are primarily in the $\hat{\phi}$ direction and dictated by the base flow $\overrightarrow{v^0}(\vec{x}, t)$ imposed by the driving mechanism, as is evident by the vertical gradient of these vectors. (Note that $\overrightarrow{v^0}$ is strictly in the $\phi z$-plane at every instant, with its magnitude linearly proportional to the height above the base plate.) In contrast, Fig.4(b) displays the sudden "jumps" of particles upon a quaking event, showing extraordinarily large displacements in apparently random directions.

In further analyses shown in the upcoming subsections, we have removed the contribution from the base flow, by defining the *fluctuating displacement* $\delta \vec{r}$ for each particle, specifically, as

$$\delta \vec{r} = \vec{r}(t) - \vec{r}(t - \delta t) - \overrightarrow{v^0}(\vec{r}, t) \delta t$$

where $\vec{r}(t)$ is the particle position at time $t$.

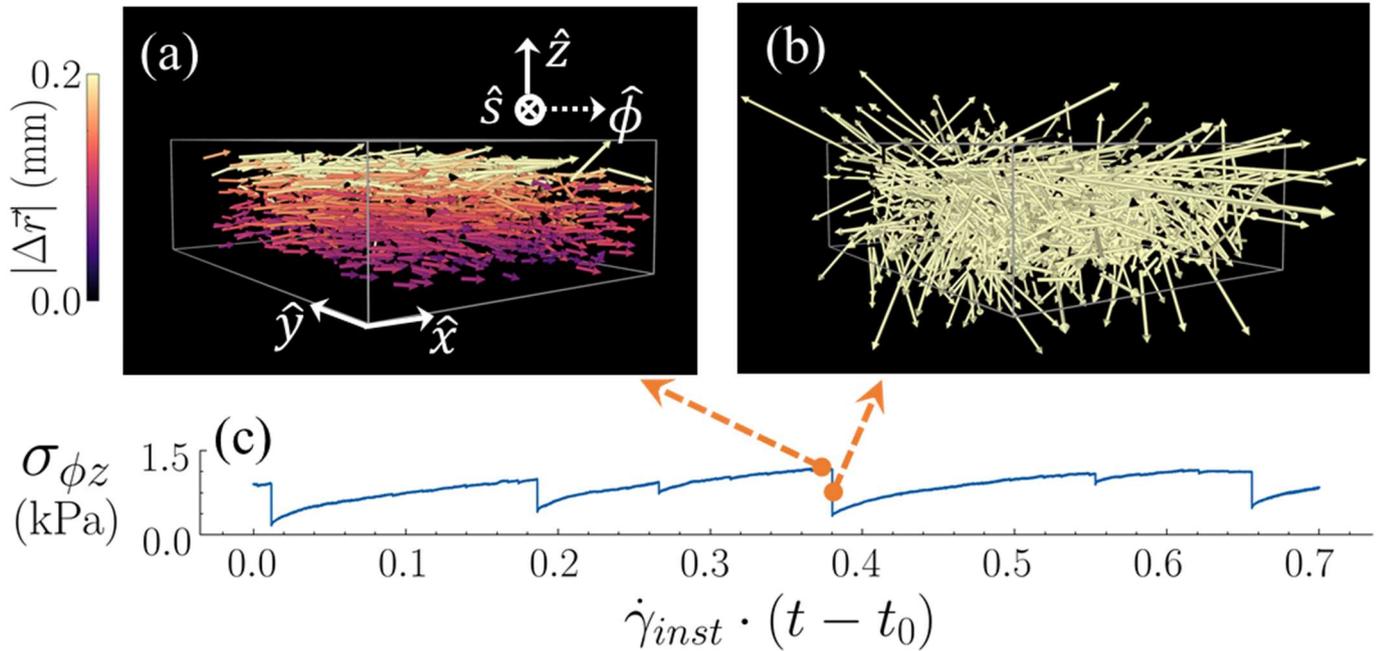

**FIG. 4** – 3D reconstructed displacements associated with two adjacent time intervals in case III, as are viewed from the $\hat{s}$ direction perpendicular to the $\phi z$-plane: **(a)** a quiescent interval immediately before a quaking event, showing the contribution from the base flow. **(b)** an interval in which a quaking occurs. **(c)** The time sequence of the instantaneous shear stress. The two intervals associated with (a) and (b) share the same effective strain step $\Delta \gamma = 3.5 \times 10^{-3}$ and the same color bar for the magnitude of displacement. For visibility, the displacement vectors in (a) and (b) are exaggerated by a factor of 100 and 20, respectively.

---

[1] Due to limited laser power and fluorescence, camera frame rate could not exceed 1000/s for a properly illuminated image. Under this constraint, case I (with $\dot{\gamma}_{inst} = 1.1 \times 10^{-1}$ s$^{-1}$) is scanned at strain intervals of $16\Delta\gamma$, while the case with $\dot{\gamma}_{inst} = 2.8 \times 10^{-2}$ s$^{-1}$ at $4\Delta\gamma$.

### A. Fluctuation over time

The rate-dependent quaking transition, outlined by the force measurements in the previous section, can now be characterized from the perspective of grain displacements. In Fig.5(a), each curve represents the cumulative distribution function of $|\delta\vec{r}|$ of all particles for one time interval. For each case, data of all time intervals are plotted. The spreading of the curves thus represents the time fluctuation of $|\delta\vec{r}|$ distribution for the granular system. The horizontal axis is scaled by the all-time average of each case $\langle|\delta\vec{r}|\rangle$.

In case I, the spread of the curves is relatively narrow, as the grain flow is smooth and homogeneous in time. The spread of the curves over time widens as the shear rate $\dot{\gamma}_{inst}$ decreases. In case II, the medians of $|\delta\vec{r}|/\langle|\delta\vec{r}|\rangle$ at different times spread throughout a range wider than one order of magnitude, whereas in case III a handful of distinct curves stand out from the majority. These curves reveal the individual quaking moments when a significant fraction of $|\delta\vec{r}|$ values jumps above $\langle|\delta\vec{r}|\rangle$, while the majority of curves that aggregate on the left correspond to the quiescent moments.

To quantify the spread of the curves, we calculate the time variation of the medians on logarithmic scale, $\Sigma$, and plot it in Fig. 5(b) as a function of $\dot{\gamma}_{inst}$. The plot shows an intriguing non-monotonic behavior with a peak at case III: As the value of $\dot{\gamma}_{inst}$ decreases toward that of case IV, even though the extreme case exhibits a huge deviation from the main cluster of curves, its occurrence corresponds to only a single quaking event (as shown in Fig.3). The decrease in $\Sigma$ with $\dot{\gamma}_{inst}$ (from III to IV) results from the rareness of large quakes at very small shear rates.

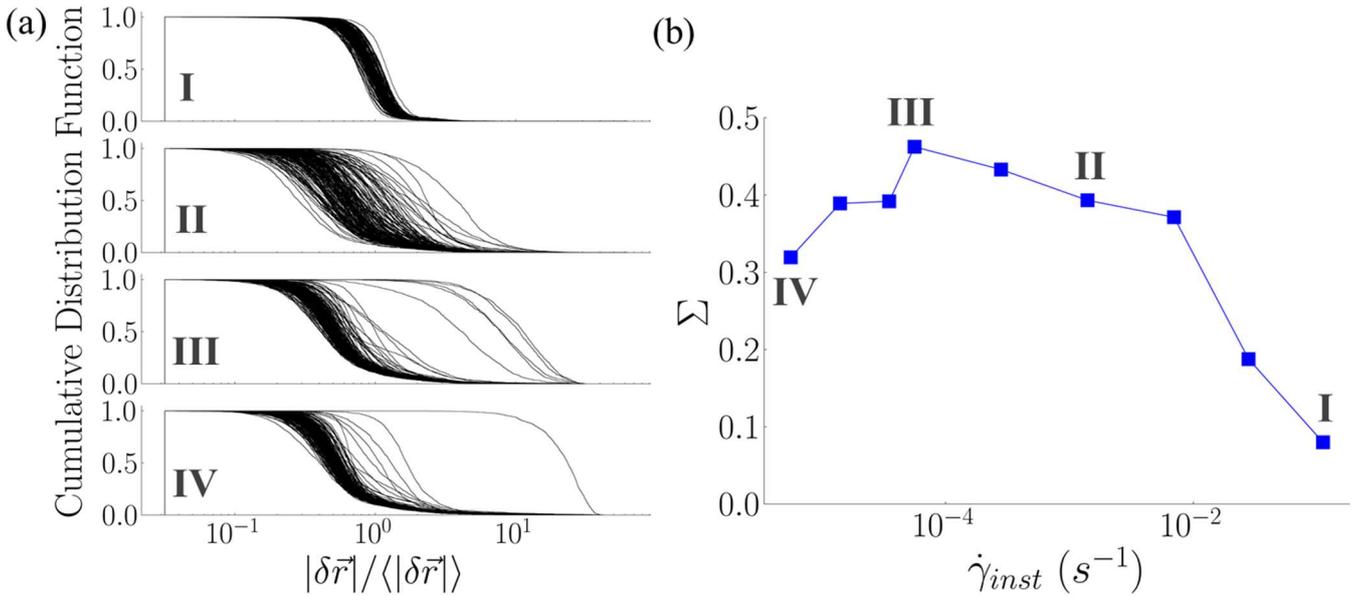

**FIG.5** Statistics based on snapshots of $|\delta\vec{r}|$. **(a)** Cumulative distribution function showing the fraction of particles that has its $|\delta\vec{r}|/\langle|\delta\vec{r}|\rangle$ above the value specified on the horizontal axis, where $\langle|\delta\vec{r}|\rangle$ stands for the average over all particles and all time for each run. **(b)** The logarithmic spread, $\Sigma$, as defined in the main text, as a function of the shear rate $\dot{\gamma}_{inst}$. The Roman numerals indicate the four signature cases defined in Fig.3.

## B.    Clusters of large displacements

We proceed to show explicitly how the rearrangements are clustered in space using a simple clustering approach. First, we find the nearest neighbors for each particle by applying Laguerre tessellation [37], which accounts for the different radii of the two types of particles. We then consider only the *significantly rearranged* particles with values of $|\delta \vec{r}|$ ranking in the top 5% in each case. (The results we obtain in this subsection are qualitatively similar for different choices of this percentile threshold -- see Appendix D.) A group of significantly rearranged particles that can be interconnected via Laguerre bonds is defined as one cluster. Following such a definition, all significantly rearranged particles can be grouped into clusters. We define the cluster size $S$ to be the number of particles in each cluster.

To understand whether these significantly rearranged particles are sparsely dispersed or tightly clustered, we plot in Fig.6(a) the occurrence of clusters of various sizes as a time sequence for each of the signature cases, in which every cluster is presented with a circle that is proportional to its size $S$. Note that the total number of particles participating the plot is approximately the same for all four cases, because experimentally the total number of tracked particles is roughly fixed.

At the fastest shear rate (case I), the values of $S$ rarely surpass 10, while clusters are scattered evenly in space and time, as illustrated in Fig. 6(b). In case II, large clusters with sizes up to several hundred emerge, which can occupy a considerable fraction of the whole volume. Interestingly, relatively large clusters in this case are often accompanied by numerous small clusters. This is visually demonstrated in the snapshot in Fig. 6(c). In case III, a handful of large clusters with $S \sim 1000$ virtually spans the entire trackable region, as shown in Fig. 6(d). These system-spanning quakes correspond to the outstanding curves in Fig. 5(a). In case IV, only one major quake remains.

Fig. 7(a) shows the all-time cumulative distributions of $S$ for each of the four cases. The results summarize the change of behaviors in a quantitative manner. We find that, from case I to III, the distribution develops a long tail that reaches size $\sim 1000$ and saturates the system, resulting in an apparent sharp cutoff near the end. The long plateau in the middle indicates a clear lack of medium-sized quakes for case III. The distribution of cluster size grows fairly extreme, especially as we go from case III to IV.

In order to highlight the effect of large clusters , we calculate the $n$-th moment of $S$, $\langle S^n \rangle$, and plot the result against $\dot{\gamma}_{inst}$ in Fig. 7(b). For $n = 1$, $\langle S \rangle$ only shows a weak dependence on $\dot{\gamma}_{inst}$ due to the fact that more than 90% of clusters are of size 10 or smaller in all cases. Higher-order moments with $n = 2$ and 4 result in a distinct peak at the middle shear rates.

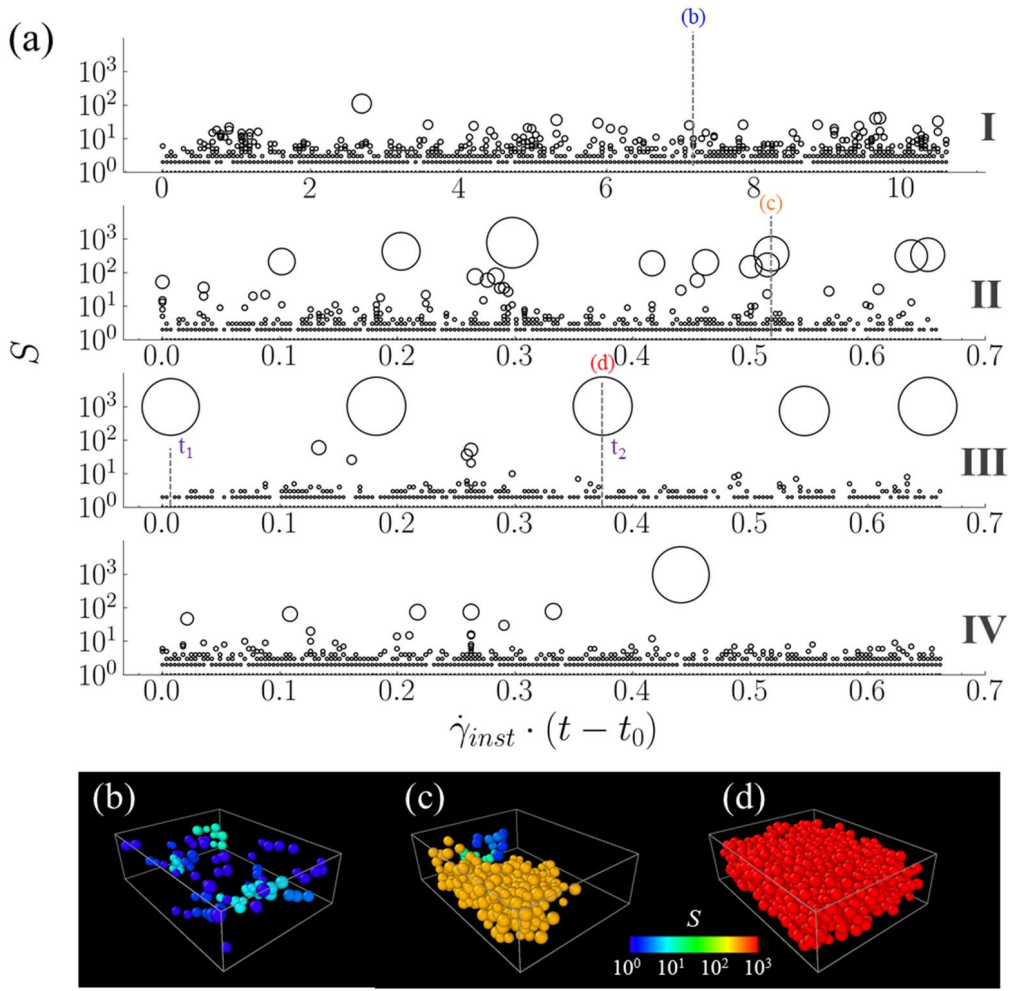

**FIG. 6 (a)** Distribution of cluster size in time for the four signature cases. Clusters are plotted as circles with area proportional to their sizes, $S$. **(b~d)** 3D reconstructed snapshots of the system, taken at specific moments marked as vertical dashed lines in (a). Two instants $t_1$ and $t_2$ are also marked for their reference in the next subsection. Only mobile particles are shown for clarity. Clusters are colored according to their size.

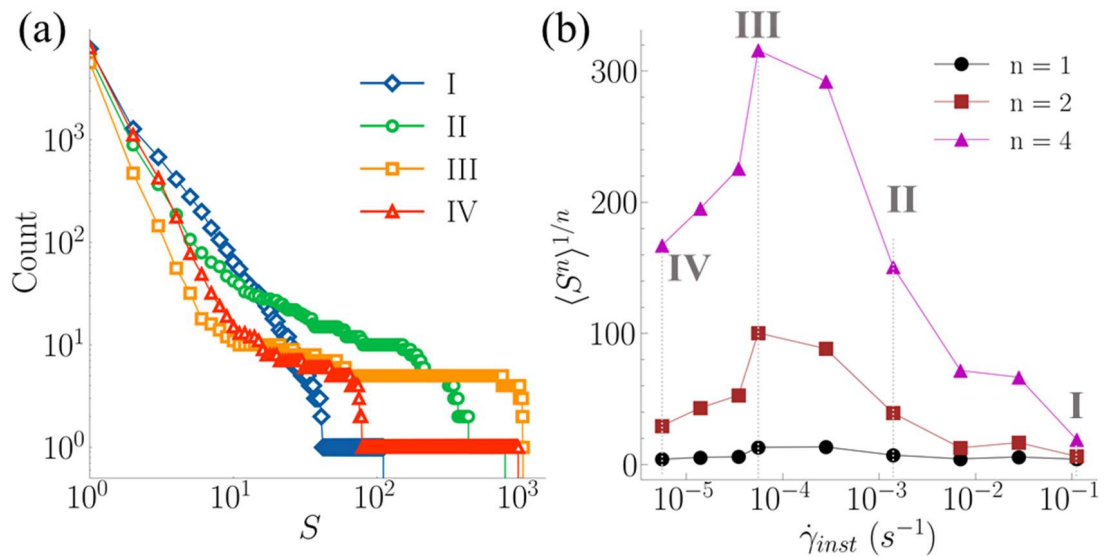

**FIG. 7 (a)** Cumulative count of clusters with size $S$ equal or larger than the number on the horizontal axis, for each of the four signature cases. **(b)** The mean value $\langle S^n \rangle^{1/n}$ with $n = 1, 2, 4$ plotted against $\dot{\gamma}_{inst}$, for all cases. The Roman numerals mark the four signature cases.

### C.  Substructure within a large cluster

The preceding analyses are all based on $|\delta \vec{r}|$, leaving the direction of motion unattended. For instance, in Fig.6, the two instants labelled as $t_1$ and $t_2$ (in case III, panel (a)) both contribute to a large mobile cluster shown as in its panel (d). However, this does not suggest that the entire cluster of particles is moving in one direction. To better understand the movements of these particles within a cluster, in this subsection we define $\delta r_\phi \equiv \delta \vec{r} \cdot \hat{\phi}$ and recover the information of direction.

In Fig.8, the values of $\delta r_\phi$ for all particles at $t_1$ and $t_2$ of case III are color-coded, both from two complementary angles of view. Panel (a) reveals that most of the particles in the upper half of the cluster move along the direction of the base flow $\hat{\phi}$ upon the quaking event at $t_1$. On the other hand, panel (b) shows that, in the quaking event at $t_2$, particles in the lower part of the cluster mostly move against the direction of $\hat{\phi}$. This reveals that there are substructures within the mobile cluster that we define in the preceding subsection, awaiting further studies in the future.

In Fig.9, we visualize the distribution of the values of $\delta r_\phi$ for all particles and for all four signature cases. This is done by color-coding small dots representing each particle according to the value of $\delta r_\phi$ and spreading them along the z-direction at every instant. Simultaneously, we display the time sequence of normal stress for each case. We see that all those heavily colored instants correspond to significant quaking events. However, not every quaking event finds its counterpart in the response of $\delta r_\phi$. This is because our current imaging analysis is limited to just half of the cell, therefore some of the localized quaking events that reside in the other half of the cell cannot be captured by the response of $\delta \vec{r}$. We are still working on improving the optics to resolve such an issue.

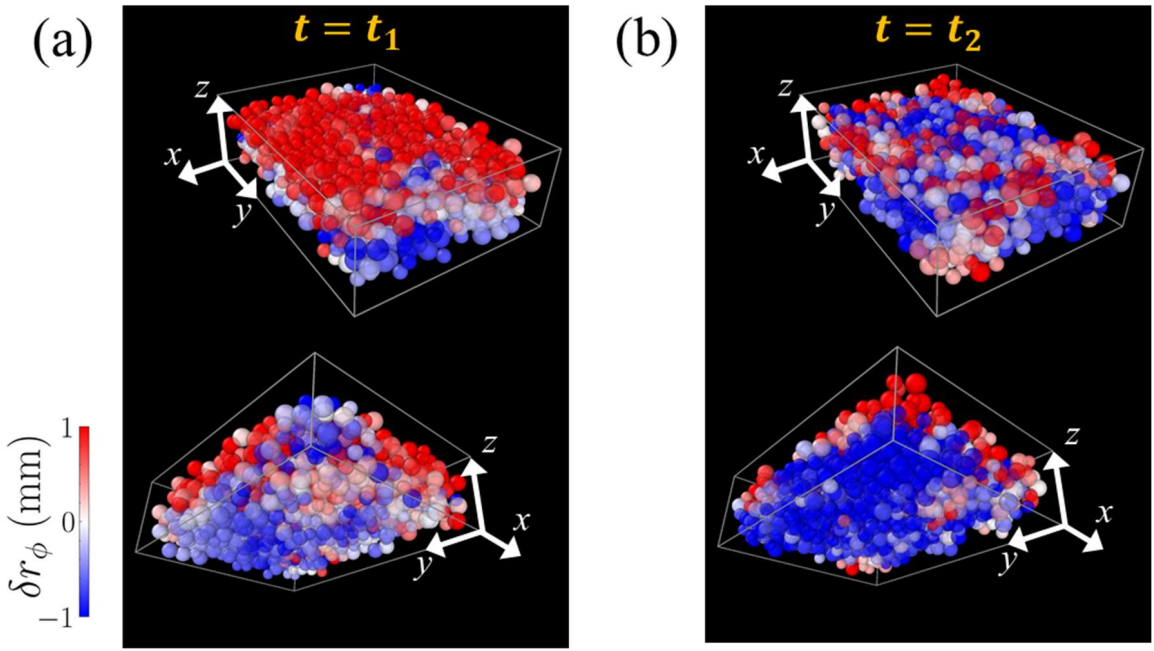

**FIG. 8**  Perspective drawings for two quaking events at **(a)** $t = t_1$ and **(b)** $t = t_2$ in case III, as indicated in Fig.6a (and Fig.9 as well). Particles are colored-coded by the value of $\delta r_\phi$ defined in the main text.  For clarity, both cases are captured from two complementary angles of view.

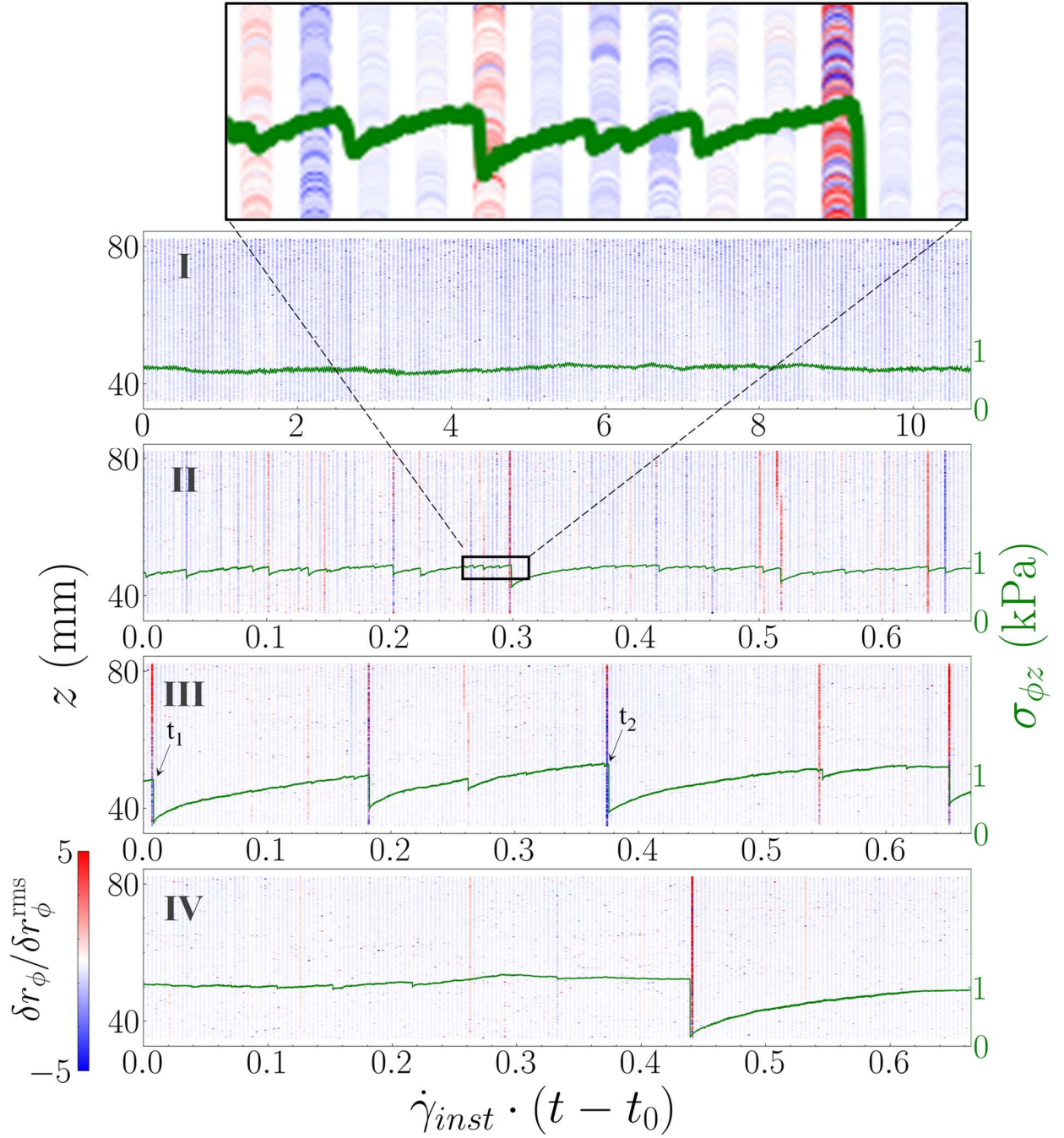

**FIG. 9** Distribution of the values of $\delta r_\phi$ of every particle, spread out along their $z$-positions (as shown by the inset on top) at every instant, overlaid by signals of the instantaneous shear stress $\sigma_{\phi z}$ (in green, with its axis on the right). Each particle is represented by a small dot that is color-coded by the value of $\delta r_\phi$ normalized by its root-mean-squared average $\delta r_\phi^{rms}$ over all particles at all times for each case. The two instants $t_1$ and $t_2$ referred to in Fig.8 are also indicated.

## IV. DISCUSSION --- Consistency with our previous work

In this section, we compare the transition identified by the present study to that reported in Ref. [17]. One common observable in the two studies is the shear stress, and the quaking transition can be described by monitoring the large drop (LD) count, defined as the number of drops in shear stress with their magnitude larger than the long-time RMS fluctuation. Fig. 10(a) displays the glaring discrepancy in LD count per strain[2] versus shear rate $\dot{\gamma}$ for the two studies, implying that the imposed shear rate is not the sole factor in this transition.

To better understand the underlying transition, we revisit the mechanism proposed in Ref. [17] that emphasizes the role of particle friction and speed-dependent lubrication. In an independent experiment [23], we have performed tribology measurements between contacting PDMS particles immersed in glycerol solution to mimic the conditions of our granular system, and found that the inter-particle friction decreases drastically once the sliding speed goes beyond a certain threshold, $V_c$, that is material dependent. This can be understood as a transition from the regime of solid-solid contact to that of mixed lubrication, commonly understood as part of the Stribeck curve that has been well established in tribology [22,23]. In our follow-up study [24], we have introduced the dimensionless shear rate $S_\ell \equiv \ell\dot{\gamma}/V_c$, which compares the generic relative speed between particles of characteristic size $\ell$ in a uniform shear flow at a rate $\dot{\gamma}$, to the threshold speed $V_c$ between contacting particles. Given the binary mixture of 9 mm and 12 mm particles being used in the current study, we set $\ell = 10$ mm as the characteristic size in calculating the dimensionless shear rate $S_\ell$. (In our previous study, we used monodisperse particles so that $\ell = d = 9$ mm.) Also by independent experiments (see Appendix E), we have found that the threshold speed $V_c$ in our current system is around 0.2 mm/s, which is substantially lower than the 5 mm/s in Ref. [17].

In Fig. 10(b), we re-plot the LD count as a function of $S_\ell$ instead of the physical shear rate, for both the previous and the current work. The result shows a reasonably good collapse of the two curves obtained from two experiments using different geometries.

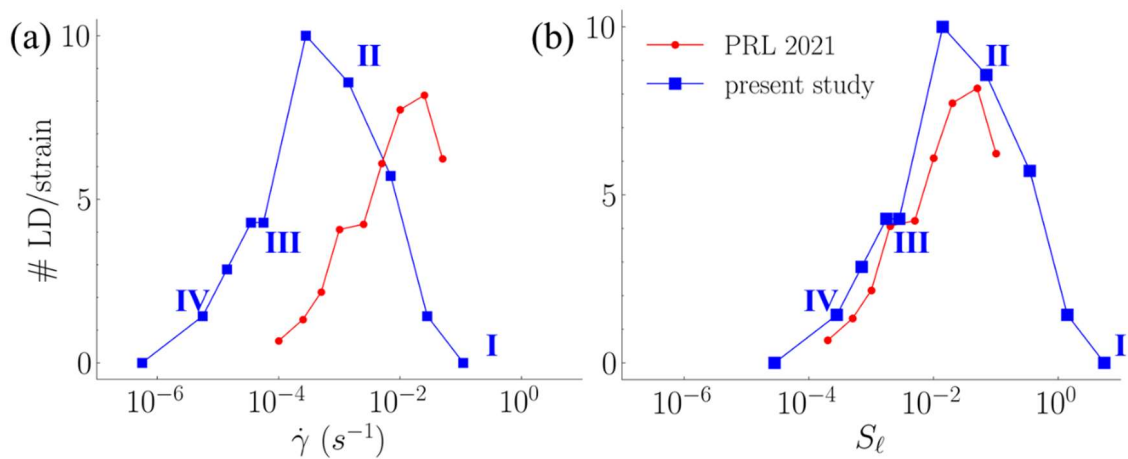

**FIG. 10**  Counts of large stress drops (LD) per unit strain plotted as a function of **(a)** the physical shear rate $\dot{\gamma}$ and **(b)** the dimensionless shear rate $S_\ell$ for data from two different systems: Our previous work published in PRL 2021 [17] and the present study in CSC setup, respectively. Roman numerals mark the four signature cases in our present study.

---

[2]  For the present study, we use the effective shear strain defined in Section II.

## V. SUMMARY

We have prototyped an experimental system in which a continuous circular shearing is imposed to soft granular particles that are tightly packed at a fixed volume. The system allows us to monitor the fluctuation of shear stress and to track the motion of the internal particles. By monitoring the sudden drops of stress, which we call quaking events, and how their occurrence changes with the shear rate, we confirm that the phenomena are consistent with our prior findings as a friction-induced transition, and that the previously proposed dimensionless shear rate ($S_\ell$) incorporates the tribology (the speed dependence of friction) involved and is indeed a good parameter to characterize such transition.

To understand how internal particles move and contribute to the quaking, we compute the fluctuating displacement $\delta\vec{r}$, in which the contribution from the base flow imposed by the driving mechanism is removed. We find that

1. The snapshots of $|\delta\vec{r}|$ reveal the trend that, at high driving rates ($S_\ell > 1$), the particles flow smoothly and the movements exhibit no significant fluctuations over time. At slower driving rates, the distribution of $|\delta\vec{r}|$ shows a bias toward small displacements, with scarce but very large displacements upon large quaking events. However, as the driving rate keeps decreasing ($S_\ell \sim 10^{-4}$), the quaking behavior becomes so extreme that only one large event occurs within our observation window (with an effective strain ~0.7). Further decrease in the driving rate renders the quaking practically undetectable.

2. By adopting only the particles with the top 5% of $|\delta\vec{r}|$ and connecting the nearest neighbors, we define clusters of significant displacement and their size $S$. The distribution of $S$ further supports the trend described above. The non-monotonic change of the mean size $<S^n>^{1/n}$ with the driving rate also reflects the trend of growing cluster size with decreasing occurrence of large clusters as the shear rate goes down.

3. By projecting $\delta\vec{r}$ onto the direction of the base flow ($\hat{\phi}$), substructures in a large cluster reflect localized but somewhat collective movements, awaiting further studies.

The current study is limited to just one packing fraction (0.59). With ongoing improvements to the system reported here, we anticipate exploring the behaviors at variable packing fractions and reporting the results in future works.

## Appendix A ----- Non-periodic Rearrangements Induced by Cyclic Driving

The cyclic movement of the cell walls implies that the imposed displacement field goes to zero after one full rotation, and particles would return to their initial positions if there were no rearrangements. However, we claim that the angle $\alpha$ is designed to be sufficiently large ($\tan \alpha = 0.45$) such that the particle configuration is incrementally rearranged. This is supported by Fig. S01, which shows the vertical trajectories of four particles that started approximately at the same height. None of the particles return to the starting point after a full rotation, suggesting that the packing has rearranged notably despite that the movement of the cell boundary is cyclic.

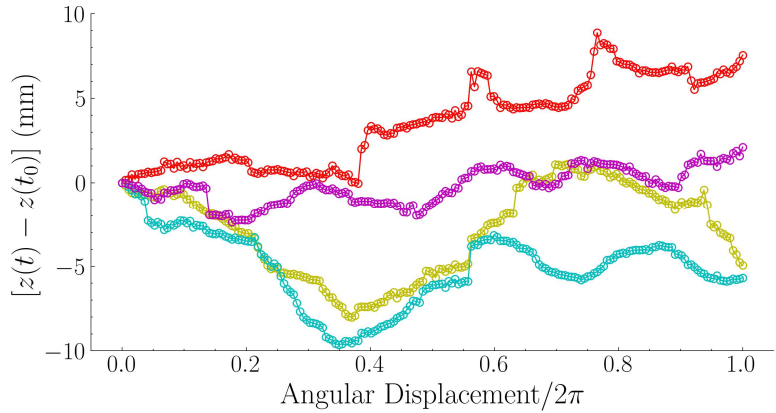

**Fig. S01** The z-trajectories of the 4 selected particles over a complete rotation, shifted by initial position $z(t_0)$. Particles are sheared at $\dot{\gamma}_{inst} = 7 \times 10^{-3}$ s$^{-1}$, and sampled at a strain interval corresponding to 1/200 rotation.

## Appendix B ----- 3D Image Processing & Particle Tracking

### 1. Effect of finite imaging distance

Experimental images are captured by a high-speed camera placed beneath the cell, while a horizontal laser sheet scans vertically. The lens (NIKON AF Nikkor 24mm F2.8) focuses at the mid-height of the cell, but there is enough depth of view for particles at both the top and the bottom of the cell to be recognizable, as shown in Fig. S02. Note that the particles appear larger in Fig. S02(a), because the imaging plane is closer to the camera than that in Fig.S02(b).

We have verified the pixel-space calibration for each slice varies fairly linearly with its distance to the camera. Based on this linear relation, we resize all image slices so that their XY-scales are all consistent with that of the lowest slice. This is done by the OPENCV method cv2.resize(interpolation=linear). Pixels that exceed the standard field of view (defined by the lowest slice) are discarded. After this process, pixel sizes for all image slices are uniform and equal to $\ell_{xy} = 0.21$ mm.

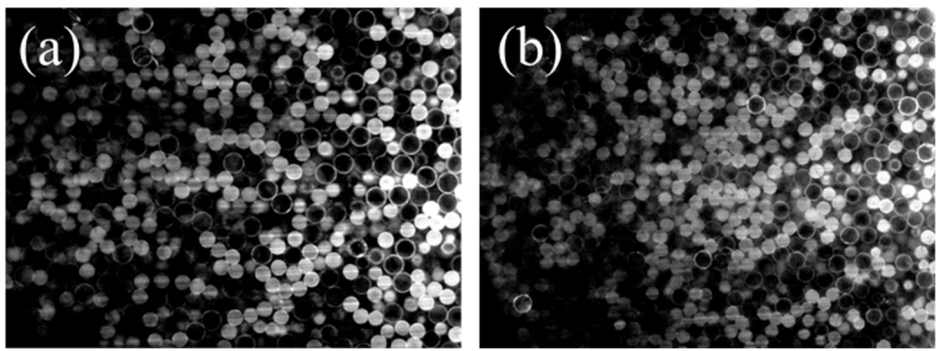

**Fig. S02** Raw images at (a) the lowest imaging plane and (b) the highest plane.

## 2. Effect of the nonlinearity of the deflection distance with respect to the rotating angle

We use a cubical water tank to deflect the laser sheet in order to achieve vertical scanning, as shown schematically in Fig. 1 in the main text. The implementation of a rotating cube naturally introduces an additional complication: The cube is rotating at a constant speed but the height of the LASER sheet as a function of time is not linear.

Fortunately, this optical problem is analytically well-defined. In our experimental setup, the height of the incident laser coincides with that of the rotating axis, as shown in Fig. S03(a). Experimentally, we have carefully measured the deflected height $\Delta h$ at different angles $\theta$ and show the results in Fig. S03(b). Meanwhile, according to Snell's law, $\Delta h$ can be derived to be:

$$\Delta h = -W \sin\theta \left[ 1 - \frac{1}{n} \cos\theta \left( 1 - n^{-2} \sin^2\theta \right)^{-\frac{1}{2}} \right]$$

, where $W = 130$ mm is the width of the cube, and $n = 1.33$ the refraction index of water. (Here, we have ignored the thickness of the glass container, which is about 5mm). Fig. A2(b) shows that the measured results agree reasonably well with the prediction of Snell's law. This curve is used in converting the "layer number" to the physical coordinate ($z$).

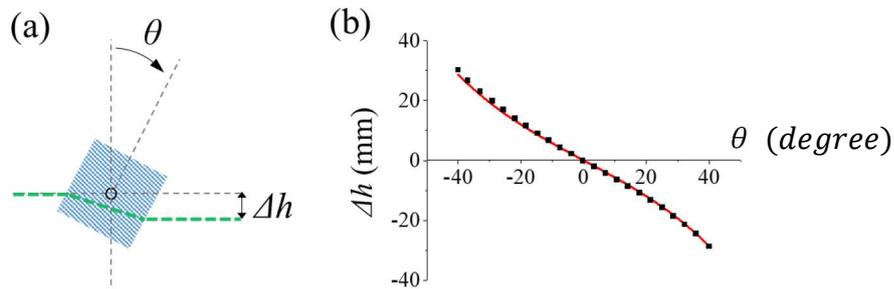

**Fig. S03** (a) Schematic drawing of a rotating cube and the deflection of LASER. The LASER enters the cube from the left. (b) Deflected height $\Delta h$ versus $\theta$. The red solid line represents the prediction from Snell's law. Black squares show the results from our measurement.

3. *Flow chart of image processing and particle tracking*

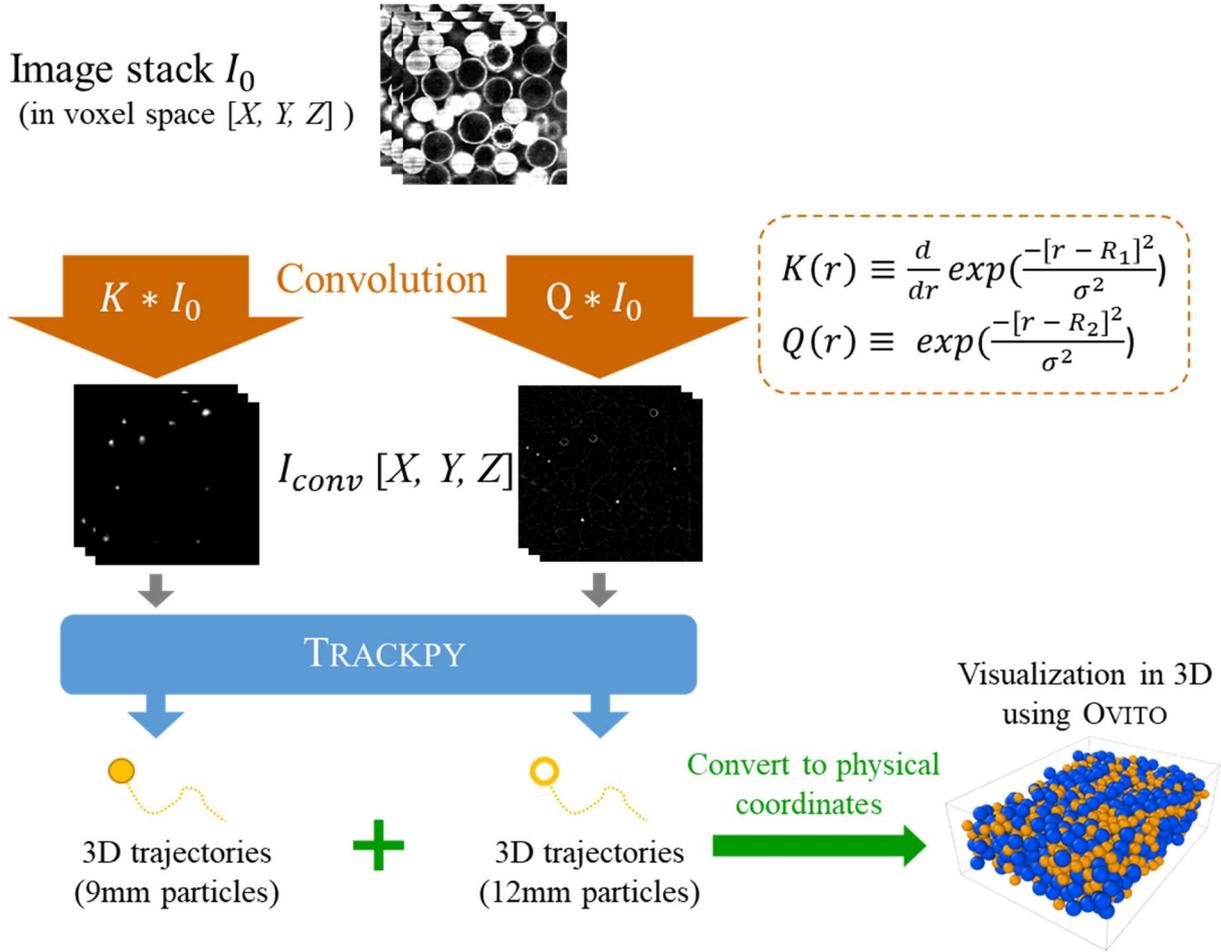

**Fig. S04**: Flow chart of image processing and particle tracking

   Our imaging processing and particle tracking routines are summarized as a flow chart (Fig.S04) above. Note that we apply two different kernels, $K(r)$ and $Q(r)$, for convolution to extract two types of particles ("filled spheres" with radius $R_1 = 9mm/2$ and "hollow spheres" with radius $R_2 = 12mm/2$), respectively, in 3D space. Also note that for a kernel centered at $[X_0, Y_0, Z_0]$ in the voxel space, the radial distance $r$ from the center to any point $[X, Y, Z]$ is defined as:

$$r(Z_0) \equiv \sqrt{\ell_{xy}^2(X - X_0)^2 + \ell_{xy}^2(Y - Y_0)^2 + \ell_z(Z_0)^2(Z - Z_0)^2}$$

, where $\ell_{xy} = 0.21$ mm is the standardized pixel size ---see Section B1, and $\ell_z(Z_0)$ is the layer distance at the height $Z_0$. The layer distance $\ell_z$ depends on the vertical position $Z_0$ because of the nonlinearity associated with Snell's Law – see Section B2. The values of $\frac{\ell_z(Z_0)}{\ell_{xy}}$ in our experiment fall in the range of 3 to 4. The adjustable parameter $\sigma$ is set to $3\ell_{xy}$.

   We then use the established algorithm TRACKPY [33,34] to locate bright features from the resulting image $I_{conv}$. Tracked coordinates in voxel space are finally converted back to physical coordinates using the standardized pixel size ($\ell_{xy} = 0.21$ mm) and the nonlinear mapping described in Section B2. In many cases, we also export our results as a format with which we can take advantage of OVITO [35,36] for the 3D visualization.

**Appendix C ----- Processing of force signals**

*1. Determination of shear force and torque*

The readings from the force sensors $s_1 \sim s_4$ are converted to forces $f_1 \sim f_4$. To determine the net force and torque on the base plate, we use $\overrightarrow{f_{xy}} = (f_2 - f_4, \; f_1 - f_3)$, and $\tau_z = (f_1 + f_3 - f_2 - f_4) \times r_{off}$, in which $r_{off}$ is defined in Fig.1.

*2. Removal of the artifact due to the imperfection of our height constraint*

In determining $f_\phi$, we find an embedded angular dependence. Fig. S05(a) shows one such example for ten consecutive cycles. Our investigation reveals that it is due to the finite rigidity of the height constraint (made of acrylic) which deforms slightly in response to the cyclic driving and creates a periodic variation of the total volume. We have resolved such an issue in our new version of circular shearing setup (to be reported in upcoming publications). Therefore we conclude that such angular dependence is indeed a mechanical artifact that does not affect our qualitative observation of quaking events. In the main texts, we have removed this artifact in presenting the time sequences of $f_\phi$:

Fig.S05(b-g) show what we have obtained as the angle-dependent average $\langle f_\phi^{Raw} \rangle$ up to ten cycles of driving, subtracted by the time-averaged value $\overline{f_\phi^{Raw}}$, from the data of six different driving rates. Despite the influence of quaking events at slow driving rates, we can see that the overall angular dependence is not very sensitive to the driving rate. Therefore we take the signal shown in Fig.S05(c) as the standard "background signal", $\langle f_\phi^{Bkgd} \rangle - \overline{f_\phi^{Bkgd}}$. For all cases shown in Fig.2(b) and other places in our main texts, $f_\phi$ has been redefined as $f_\phi \equiv f_\phi^{Raw} - ( \langle f_\phi^{Bkgd} \rangle - \overline{f_\phi^{Bkgd}} )$, where the raw signal $f_\phi^{Raw}$ is always subtracted by the standard background.

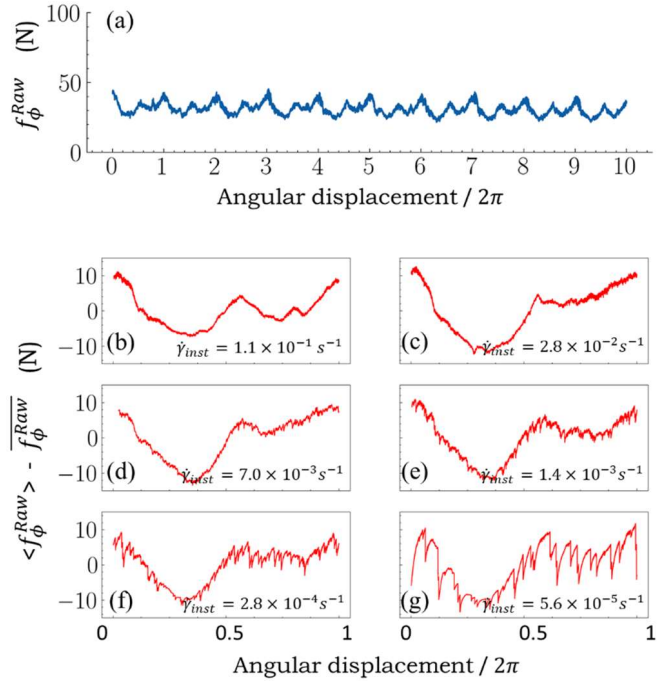

**Fig. S05:** (a) The raw signal obtained from ten consecutive rotation cycles at $\dot{\gamma}_{inst} = 1.12 \times 10^{-1}$ s$^{-1}$, plotted against the accumulated angular displacement. (b~g) The angle-dependent average subtracted by its time-averaged $\langle f_\phi^{Raw} \rangle - \overline{f_\phi^{Raw}}$, for six different values of $\dot{\gamma}_{inst}$ in decreasing order.

## Appendix D ----- Effect of the threshold percentile

The definition of cluster involves a threshold percentile, $\xi_p$, used to separate particles into two groups: those that significantly rearrange, and those that do not. The results regarding clusters presented in the main text would thus depend explicitly on the arbitrary choice of $\xi_p$. We use $\xi = 5\%$ in the main texts. To verify the robustness, we perform the same clustering analysis using $\xi_p = 2\%$ and $\xi_p = 10\%$, and show the results in Fig. S06. For all four states, lowering $\xi_p$ appears to boost the number of small clusters and enlarge major ones, but the overall trend, that is, the rate dependence of the clustering, remains qualitatively similar.

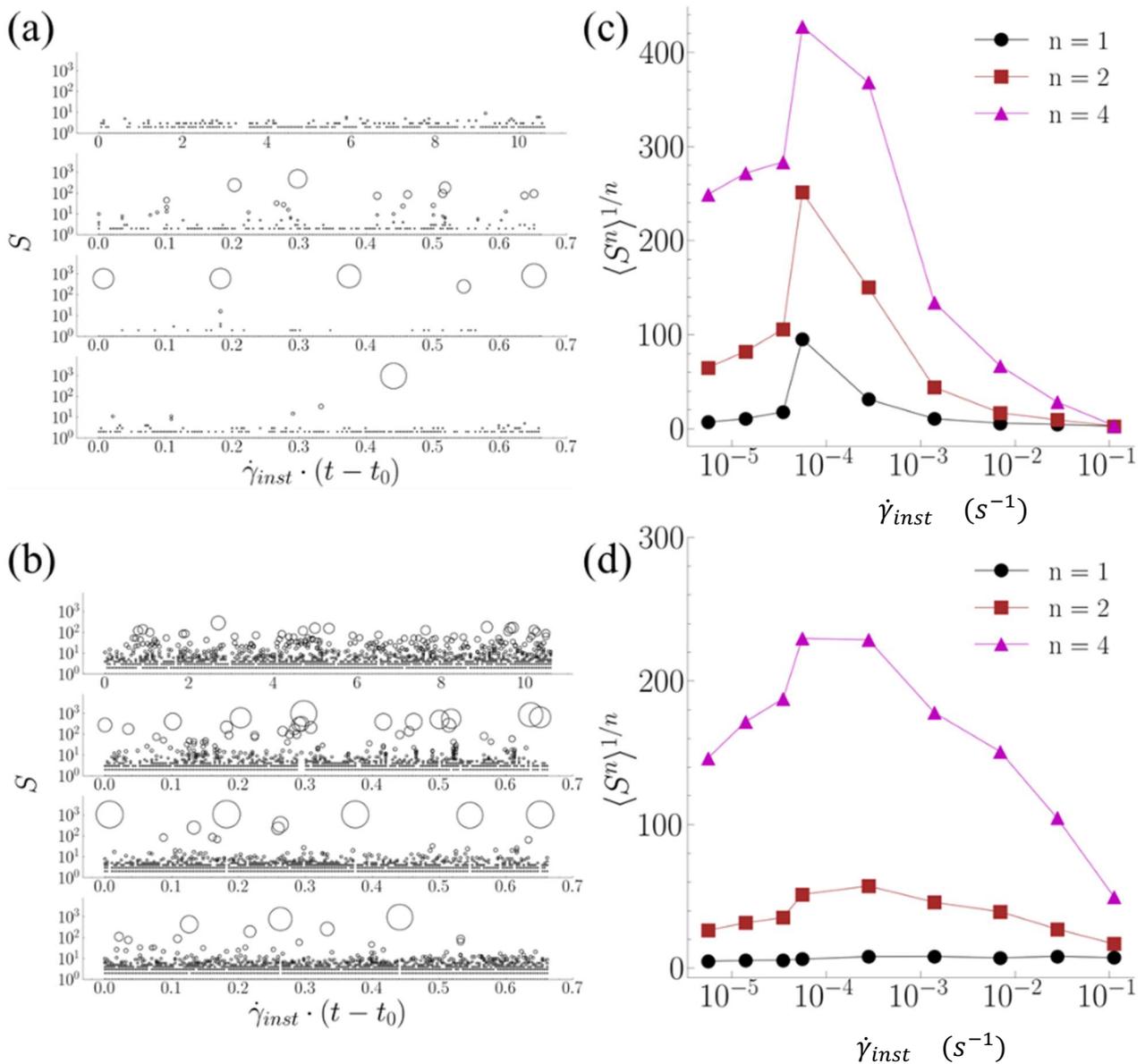

**Fig. S06:** Counterparts of Fig. 6 and Fig.7(b) in the main texts using two different values of $\xi_p$. Results in panel (a) and (c) are computed for $\xi_p = 2\%$, while those in (b) and (d) are for $\xi_p = 10\%$.

**Appendix E ----- Long-time change of the PDMS surface**

For years, we have been monitoring the gradual decline of the characteristic sliding speed, $V_c$, for the PDMS particles involved in our experiments, using the same device and methodology as we reported in Ref. [23]. PDMS is porous. It is known that submerging PDMS in oil-based liquid can change its tribology substantially---see Ref.[38] for an example. Freshly molded PDMS particles (two years ago) had $V_c \sim 5$ mm/s. After submerging in the glycerol-water mixture for a year, the value of $V_c$ for these PDMS particles was determined to be 1 mm/s. Immediately after all data reported in this paper were collected, the value of $V_c$ was determined to be 0.2 mm/s, based on which we calculate our dimensionless shear rate $S_\ell$ in Section IV.